\documentclass[12pt,preprint]{aastex}

\newcommand{\kms}{km s$^{-1}$}
\newcommand{\Bband}{B}
\newcommand{\Vband}{V}
\newcommand{\BminusV}{({\Bband}{\rm -}{\Vband})}

\newcommand{\Ebv}{E\BminusV}

\newcommand{\halpha}{H$\alpha$}

\newcommand{\ssp}{\def\baselinestretch{1.0}\large\normalsize}
\newcommand{\gtrsi}{\mathrel{\hbox{\rlap{\hbox{\lower4pt\hbox{$\sim$}}}\hbox{$>$}}}}

\shorttitle{SN 2002ap} \shortauthors{Leonard et al.}

\begin{document}

\title{Photospheric-Phase Spectropolarimetry and Nebular-Phase Spectroscopy of
the Peculiar Type Ic Supernova 2002ap}

\vspace{2cm}

\author{Douglas C. Leonard,}
\affil{Five College Astronomy Department, University of Massachusetts, Amherst,
 MA 01003-9305; leonard@nova.astro.umass.edu}
\author{Alexei V. Filippenko, Ryan Chornock, and Ryan J. Foley}
\affil{Department of Astronomy, University of California, Berkeley,
 CA 94720-3411; (alex, chornock, rfoley)@astro.berkeley.edu}

\vspace{1cm}

\begin{abstract}

The early-time optical spectrum of the Type Ic supernova (SN) 2002ap was
characterized by unusually broad features, leading some authors to designate it
a ``hypernova." We present optical spectropolarimetry of this object 16 and 37
days after the estimated date of explosion.  After correcting for interstellar
polarization, we find evidence for a high level of intrinsic continuum
polarization at both epochs: $p \gtrsim 1.3\%$ on day 16 and $p \gtrsim 1.0\%$
on day 37. Prominent line polarization is also seen, especially in the trough
of the Ca~II near-infrared triplet during the second epoch. When interpreted in
terms of the oblate, electron-scattering model atmospheres of H\"{o}flich
(1991), our results imply an asymmetry of at least $20\%$ (day 16) and $17\%$
(day 37).  The data suggest a fair degree of axisymmetry, although the
polarization angle of the dominant polarization axes are different by about
$55^\circ$ between the two epochs, implying a complex morphology for the
thinning ejecta.  In particular, there exists some spectropolarimetric evidence
for a different distribution of Ca relative to iron-group elements.

We also present flux spectra of SN~2002ap taken 131 and 140 days after the
explosion.  The spectra are characterized by a very weak continuum and broad
emission lines, indicating that SN~2002ap has entered the nebular phase.  The
spectral features are substantially similar to those of ``normal'' SNe Ic in
the nebular phase, and the emission lines are not significantly broader.
However, some of the broad lines are very sharply peaked, and may possess a
narrow component (probably unresolved by our spectra, FWHM $\lesssim
400$ \kms) that is redshifted by $\sim 580$ \kms\ with respect to the systemic
velocity of the host galaxy. 

\end{abstract}

\medskip

\section{INTRODUCTION}
\label{sec:introduction}

Although core-collapse supernovae (SNe) present a wide range of spectral and
photometric properties, there is growing consensus that much of this variety is
due to the state of the progenitor star's hydrogen and helium envelopes at the
time of explosion. Those stars with massive, intact envelopes produce SNe of
Type II-plateau (SNe II-P; II due to the presence of H in the spectrum, and
``plateau'' due to the characteristic shape of the light curve), those that
have lost their entire hydrogen envelope (perhaps through stellar winds or mass
transfer to a companion) result in SNe~Ib (Type I because of the absence of
obvious hydrogen in their spectra), and those that have been stripped of both
hydrogen and most (or all) of their helium produce SNe~Ic; see Filippenko
(1997) for a general review.

Recently, a new subclass of objects has emerged whose members generically
resemble SNe~Ic (no hydrogen or obvious helium spectral features), but, unlike
traditional SNe~Ic, have spectra characterized by unusually broad features at
early times, indicating velocities in excess of $\sim 30,000$ \kms.  A few also
possess inferred kinetic energies exceeding that of ``normal'' core-collapse
SNe by more than a factor of 10 (see, e.g., Nomoto et al. 2001).  These objects
are colloquially referred to as ``hypernovae,'' although not all of them are
clearly more luminous or energetic than normal SNe~Ic.  There are currently 5
generally accepted members of this rare class: SN~1997dq, SN~1997ef, SN~1998bw,
and, most recently, SN 2002ap and SN~2002bl (see, e.g., Iwamoto et al. 2000;
Matheson et al. 2000b; Filippenko, Leonard, \& Moran 2002; Mazzali et al. 2002;
Kawabata et al. 2002).  A related subclass of SNe exhibits many of the
characteristics of these objects, but with hydrogen present in the spectra; the
clearest examples are SN 1997cy and SN 1999E (see Filippenko 2001, and
references therein), and they, too, are sometimes called hypernovae.

Intense interest in hypernovae has been sparked not only by their peculiar
spectral features, but also by the strong spatial and temporal association
between the brightest and most energetic of these events, SN 1998bw, and the
$\gamma$-ray burst (GRB) 980425 (e.g., Galama et al. 1998).  This potential
association has fueled the proposition that some (or, perhaps all)
core-collapse SNe explode due to the action of a ``bipolar'' jet of material
(Wheeler, Meier, \& Wilson 2002; Khokhlov et al. 1999; MacFadyen \& Woosley
1999), as opposed to the conventional neutrino-driven mechanism (Colgate \&
White 1966; Burrows et al. 2000, and references therein).  Under this paradigm,
a GRB is only produced by those few events in which the progenitor has lost all
of its outer envelope material (i.e., it is a ``bare core'' collapsing), and is
only observed if the jet is closely aligned with our line-of-sight (l-o-s).
With such an explosion mechanism, one can well imagine that severe distortions
from spherical symmetry may exist in the ejecta.

Since all hot, young SN atmospheres are dominated by electron scattering
(Wagoner 1981), which is an inherently polarizing process, spectropolarimetry
provides a powerful probe of young SN morphology.  The basic question is
whether the SN is round: by symmetry, if the atmosphere is spherical the
directional components of the electric vectors cancel to produce zero net
polarization.  In contrast, an asymmetric atmosphere will yield a nonzero
polarization due to incomplete cancellation.  From recent polarimetry of
core-collapse events, a tentative trend has been identified: the degree of
polarization (and, hence, asymmetry) increases with decreasing progenitor
envelope mass (Wheeler 2000). Indeed, while the intrinsic polarizations of
``normal'' SNe II-P are generally below $1\%$ (Leonard et al. 2001; Leonard \&
Filippenko 2001), a polarization of over 4\% has been reported for an SN Ic,
suggesting an axial asymmetry of more than 3:1 in this event (Wang et
al. 2001).  The number of events investigated in detail with
spectropolarimetry, however, remains small.

An additional technique that has been used to infer explosion asymmetry is the
analysis of nebular line profiles.  Recent simulations involving explosive
nucleosynthesis in aspherical, jet-induced SN explosions predict that
intermediate-mass and heavy elements such as iron are ejected (at high
velocity) primarily along the poles whereas elements synthesized in the
progenitor (e.g., He, C, Ca, O) are preferentially located at lower velocities
near the equatorial plane in the expanding ejecta (Maeda et al. 2002; Khokhlov
\& H\"{o}flich 2001; H\"{o}flich et al. 2001).  Maeda et al. (2002) model the
effect that explosion asymmetry has on the spectra of hypernova explosions, and
compare the results to the observed nebular line profiles of SN~1998bw.  The
main observable diagnostic for explosion asymmetry is found to be the ratio of
the width of a probable [\ion{Fe}{2}] blend near 5200~\AA\ to that of
[\ion{O}{1}] $\lambda\lambda$6300, 6364~\AA.  Maeda et al. (2002) find models
of spherical explosions to be incapable of yielding velocity ratios greater
than $\sim$ 3:2, whereas aspherical models can generate ratios of more than
2:1.  Although a specific degree of asphericity is not given, Maeda et
al. (2002) find the nebular-phase line-width ratio of [\ion{Fe}{2}] to
[\ion{O}{1}] for SN~1998bw to be inconsistent with a spherical explosion.

In this paper, we report on spectropolarimetric observations obtained for the
recent peculiar\footnote{Berger, Kulkarni, \& Chevalier (2002) recently used
radio observations to conclude that SN~2002ap does not exhibit evidence for a
large amount of relativistic ejecta, and hence is an ``ordinary" SN~Ic;
however, we feel that it remains a somewhat peculiar SN~Ic, given its unusually
broad early-time spectral features.}  Type Ic SN~2002ap in M74 (Kinugasa et al. 2002;
Filippenko \& Chornock 2002; Mazzali et al.  2002; Gal-Yam, Ofek, \& Shemmer
2002) during the photospheric phase, and two nebular-phase optical flux
spectra.  We describe the observations in \S~\ref{sec:observations}, present
our results and analysis in \S~\ref{sec:resultsandanalysis}, and discuss our
conclusions in \S~\ref{sec:conclusions}.  Additional details of the
observations and analysis are given in Appendix~A.  As this paper was nearing
completion, a submitted paper by Kawabata et al. (2002) describing similar
spectropolarimetric data appeared on astro-ph; we briefly compare our results
with that work in \S~\ref{sec:conclusions}.

\section{Observations}
\label{sec:observations}

We obtained spectropolarimetry of SN~2002ap on 2002 February 14 and March 7 (16
and 37 days, respectively, after the time of explosion derived by Mazzali et
al. 2002), with the Low-Resolution Imaging Spectrometer (Oke et al. 1995) in
polarimetry mode (LRISp; Cohen 1996\footnote{Instrument manual available at
\url{http://www2.keck.hawaii.edu:3636/}.}) at the Cassegrain focus of the
Keck-I 10-m telescope.  After spending nearly three months in solar
conjunction, SN~2002ap emerged from behind the Sun in June, and we obtained
optical flux spectra of it on 2002 June 8 and June 17 (131 and 140 days,
respectively, after the time of explosion) with the Kast double spectrograph
(Miller \& Stone 1993) at the Cassegrain focus of the Shane 3-m telescope at
Lick Observatory. On the second June epoch, spectropolarimetry of five distant
Galactic stars near to the l-o-s of SN~2002ap was also obtained in order to
quantify the Galactic interstellar polarization (ISP) component contributing to
the observed polarization of SN~2002ap.  (Due to the very limited time of
visibility, spectropolarimetry of SN~2002ap itself was not obtained on this
date.)  The slit position angle was set to within $3^{\circ}$ of the
parallactic angle (Filippenko 1982) for all observations in order to minimize
differential light loss, and the total-flux spectra from all epochs were
corrected for continuum atmospheric extinction and telluric absorption bands
(e.g., Wade \& Horne 1988; Bessell 1999).  A journal of observations is given
in Table~1.

The polarimetry data were reduced according to the methods outlined by Miller,
Robinson, \& Goodrich (1988) and detailed by Leonard et al. (2001) and Leonard
\& Filippenko (2001). Additional details of the observations and analysis,
including an investigation of the potential impact of second-order light and
instrumental polarization on our results, are given in Appendix~A.  The
polarization angle (P.A.)  offset between the half-wave plate and the sky
coordinate system was determined for all polarimetry epochs (at both Keck and
Lick) by observing the polarized standard star BD $+59^\circ389$ and setting
its $V$-band polarization position angle (i.e., $\theta_V$, the debiased,
flux-weighted average of the polarization angle over the wavelength range
5050--5950~\AA; see Leonard et al. 2001) equal to $98.09^{\circ}$, the value
cataloged by Schmidt, Elston, \& Lupie (1992).  To check the precision of the
P.A. offset at Keck, we also observed the polarized standard star HD 127769 on
both nights, and found its P.A. to agree to within $2.4^{\circ}$ of the values
measured by Clemens \& Tapia (1990) and Hiltner (1956). Since the values
individually reported by the two catalogs disagree by $1.3^\circ$ in P.A. and
$0.14\%$ in polarization level for this standard star, we do not ascribe much
significance to the disagreement found between our measurements and the
cataloged values.  It is possible that this star exhibits polarization
variability. The observed null standard star, HD 57702 (Mathewson \& Ford
1970), was found to be null to within $0.08\%$ on both nights at Keck.  (See
Appendix~A for details of the small amount of instrumental polarization that
may have existed in one of our settings during the February 14 epoch.)  At
Lick, the P.A. offset was checked by observing HD 204827; its P.A., measured to
be $58.1^\circ$, differs by only $0.6^\circ$ from the value cataloged by
Schmidt et al. (1992).  The null standards observed at Lick, BD +$32^\circ3739$
and HD 212311, were both found to be null to within $0.11\%$.


\begin{figure}
\ssp
\begin{center}
\rotatebox{0}{
\scalebox{1.0}{
\plotone{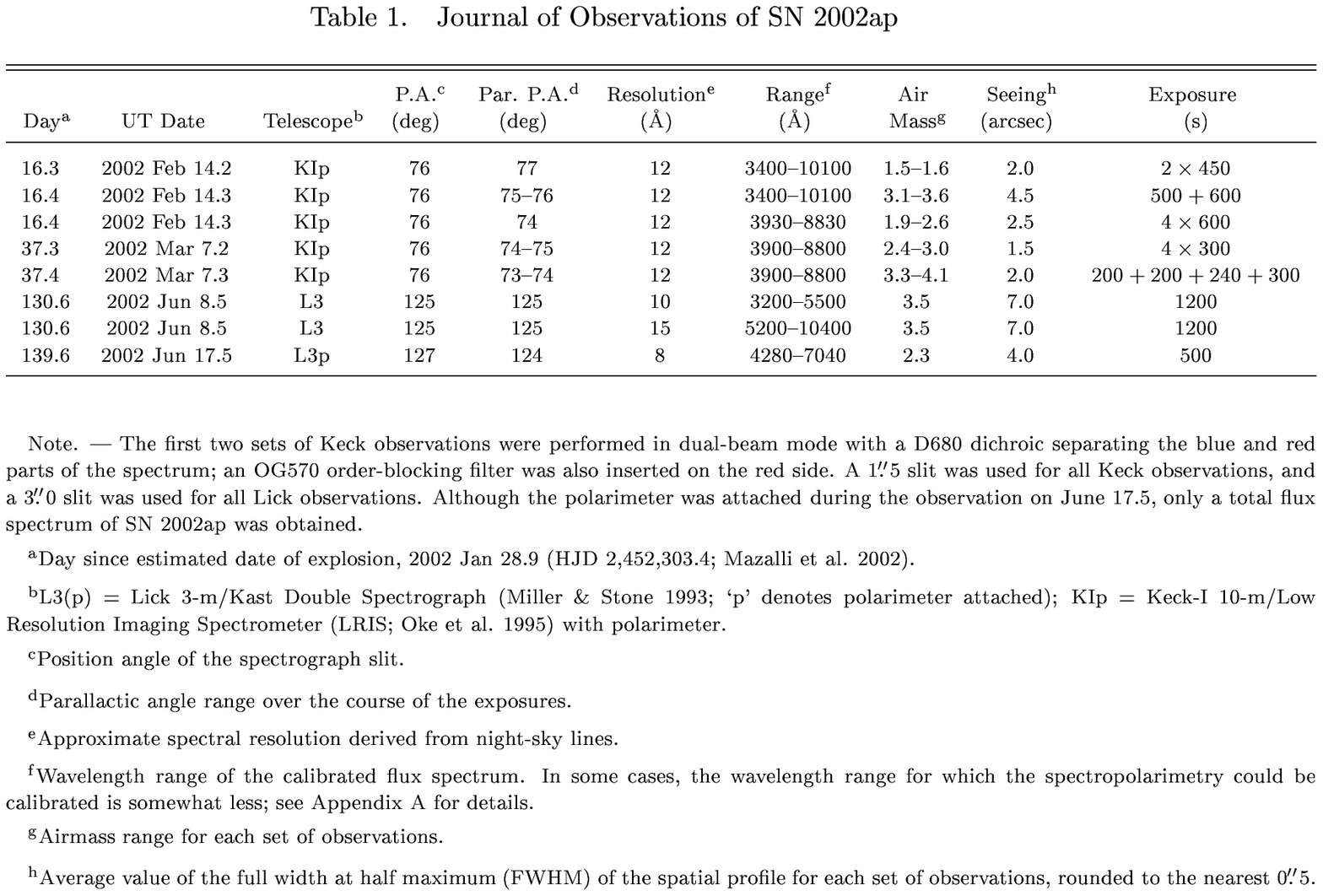}
}
}
\end{center}
\end{figure}


\begin{figure}
\ssp
\begin{center}
\rotatebox{0}{
\scalebox{0.7}{
\plotone{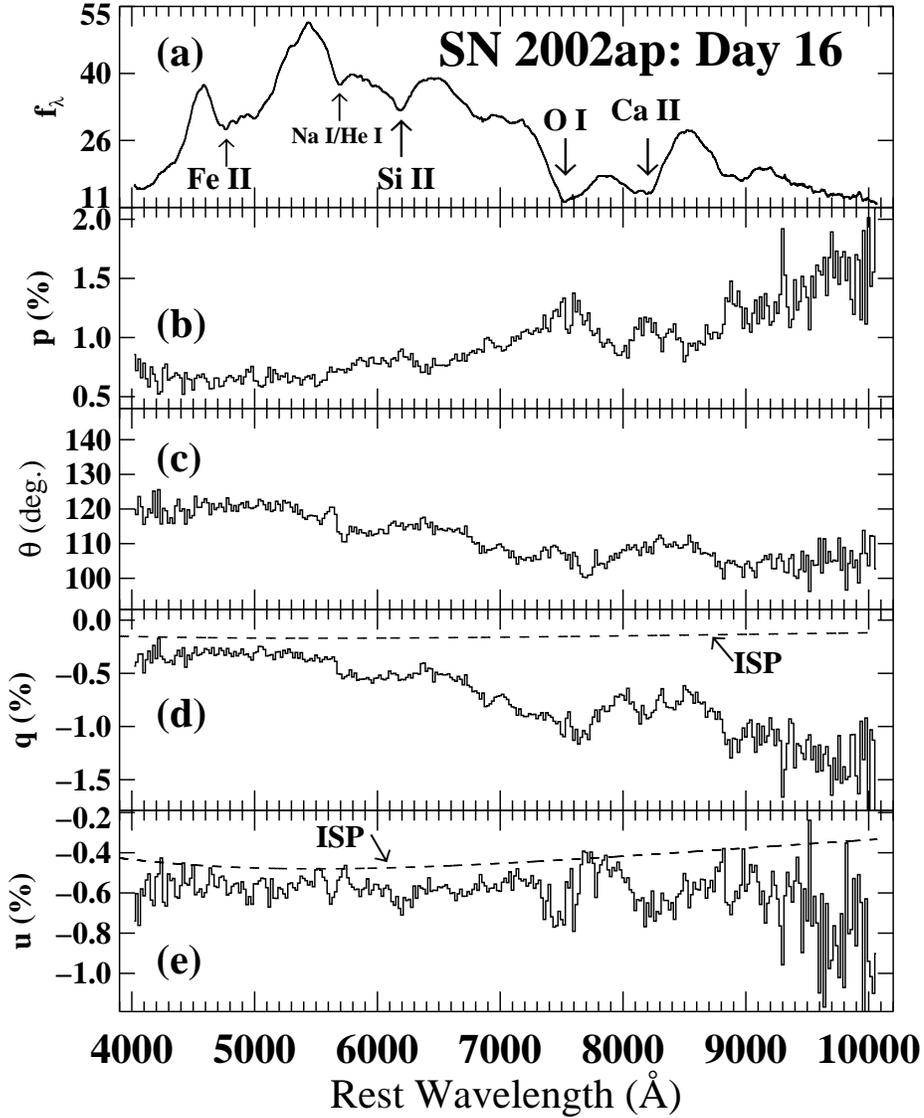}
}
}
\end{center}
\caption{Polarization data for SN~2002ap obtained 2002 February 14, about 16
days after explosion (Mazzali et al. 2002).  The NASA/IPAC Extragalactic
Database (NED) recession velocity of 657 km s$^{-1}$ for M74 has been removed
in this and all figures.  ({\it a}) Total flux, in units of $10^{-15}$ ergs
s$^{-1}$ cm$^{-2}$ \AA$^{-1}$, with prominent absorption features identified by
Mazzali et al. (2002) indicated.  ({\it b}) Observed degree of polarization.
({\it c}) Polarization angle in the plane of the sky. ({\it d, e}) The
normalized $q$ and $u$ Stokes parameters, with the level of ISP estimated in
\S~\ref{sec:removalofisp} indicated ({\it dashed lines}).  The total-flux
spectrum is shown at 2~\AA\ bin$^{-1}$, whereas the polarization data is binned
at 20~\AA\ bin$^{-1}$ to improve the signal-to-noise ratio.  Unless otherwise
indicated, the displayed polarization $p$ is the ``rotated Stokes parameter'' (RSP;
see Leonard et al. 2001) in all figures.
\label{fig:1} }
\end{figure}


\begin{figure}
\ssp
\begin{center}
\rotatebox{0}{
\scalebox{0.7}{
\plotone{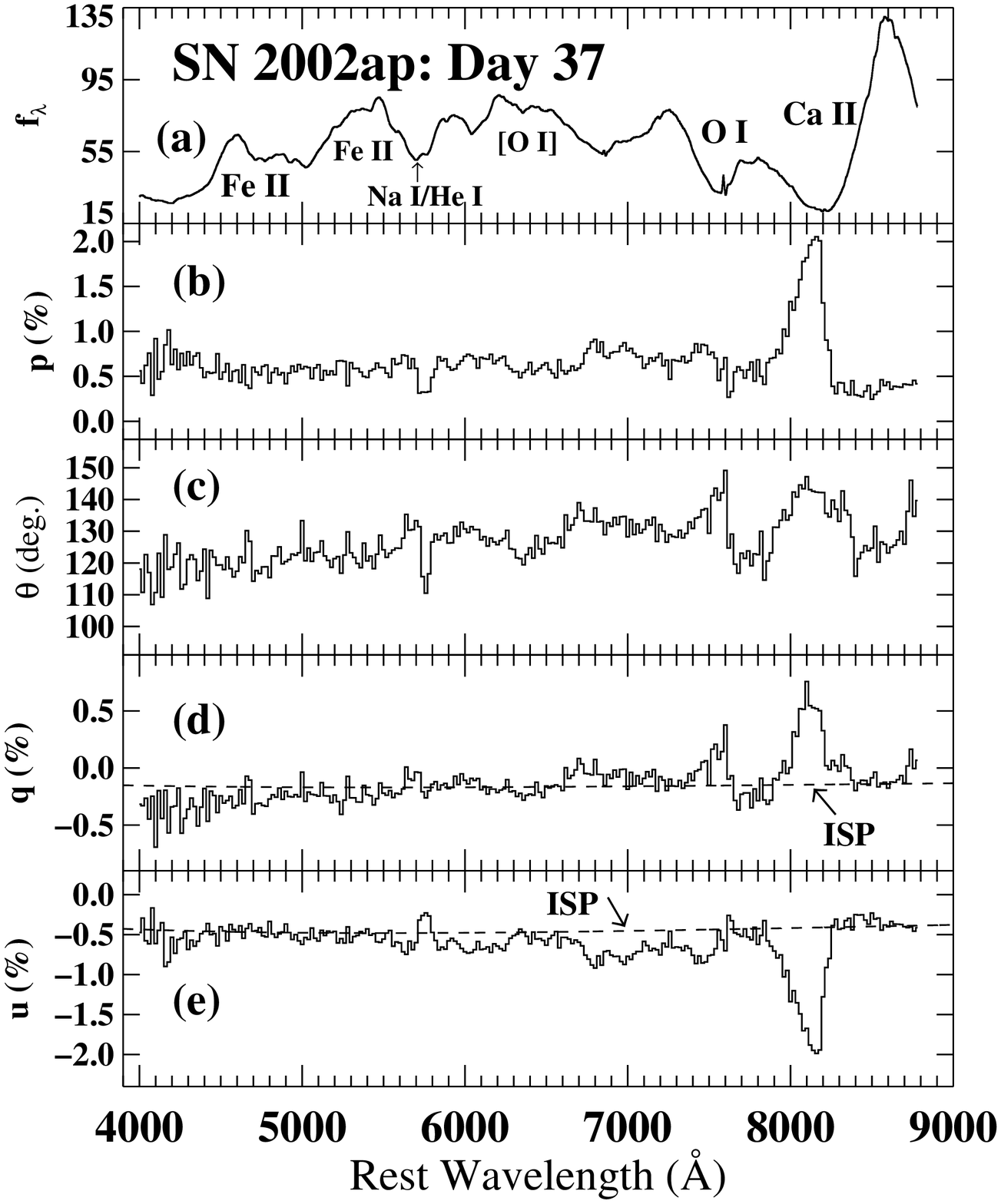}
}
}
\end{center}
\caption{As in Figure~\ref{fig:1}, but for data obtained on 2002 March 7, about
37 days after explosion (Mazzali et al. 2002).  Probable feature
identifications are taken from Mazzali et al. (2002) and Patat et al. (2001).
\label{fig:2} }
\end{figure}

\section{Results and Analysis}
\label{sec:resultsandanalysis}
\subsection{Observed Polarization}
\label{sec:observedpolarization}

The results for SN~2002ap are shown in Figures~\ref{fig:1} and \ref{fig:2}.
Before estimating and then removing the ISP (\S~\ref{sec:removalofisp}), we
make a few general remarks about the observed polarization.  The nature of the
polarization of SN~2002ap changed significantly between our two observations.
On day 16, the observed polarization rises from $\sim 0.6\%$ near 4000~\AA\ to
$\gtrsim 1.5\% $ at 10,000~\AA, with the general rise punctuated by some
features that may be associated with strong features seen in the total-flux
spectrum.  The P.A. also changes by about $15^\circ$ across the spectrum.  The
notable feature in the polarization data from day 37 is an enormous
polarization change associated with the P-Cygni profile of the \ion{Ca}{2}
near-infrared (IR) triplet.  The polarization changes between the two observational
epochs, as well as the strong modulations across spectral lines, indicate that
some of the observed polarization is intrinsic to SN~2002ap, and is not
interstellar in nature.  In order to progress with further interpretation,
however, we must attempt to estimate and remove the ISP from our data.

\subsection{Removal of ISP}
\label{sec:removalofisp}

From reddening considerations we do not expect a large ISP, since any dust
that polarizes the SN light should also redden it, and SN~2002ap is not highly
reddened: $\Ebv_{\rm MW} = 0.07$ mag (Schlegel, Finkbeiner, \& Davis 1998) and
$\Ebv_{\rm M74} = 0.01$ mag (Klose, Guenther, \& Woitas 2002), for a total
estimated reddening of $\Ebv \approx 0.08$ mag.  Since the empirical Galactic
limit for the maximum observed polarization for a given reddening is ${\rm
ISP}/{\Ebv} < 9.0\%\ {\rm mag}^{-1}$ (Serkowski, Mathewson, \& Ford 1975; see,
however, Leonard et al. 2002a for a possible violation of this inequality for
dust in an external galaxy), this implies an upper limit to the total (Galactic
plus host galaxy) ISP of only $0.72\%$.


\begin{figure}
\ssp
\begin{center}
\rotatebox{0}{
\scalebox{0.7}{
\plotone{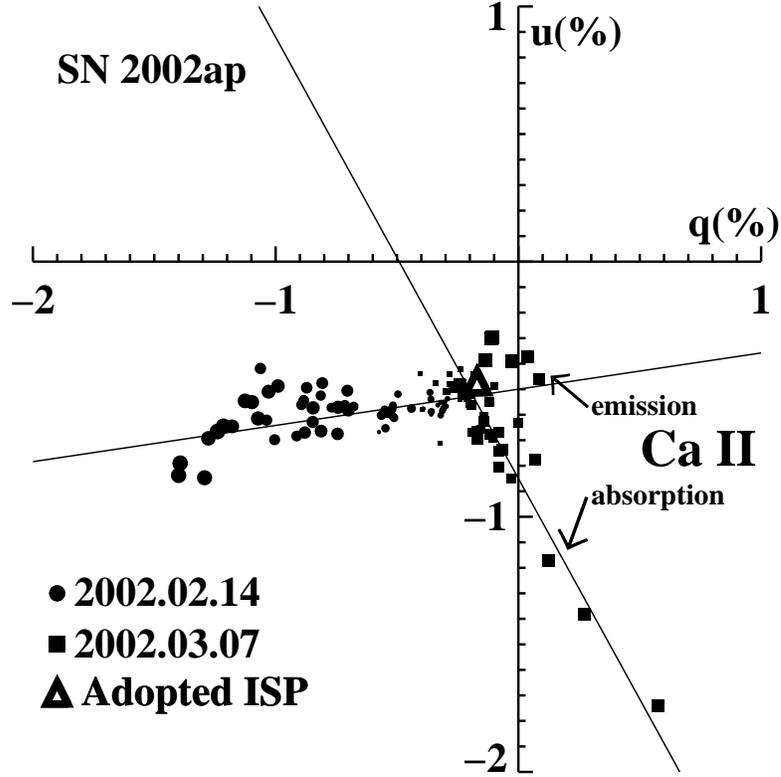}
}
}
\end{center}
\caption{Polarization data in the $q$-$u$ plane for SN 2002ap on 2002 Feb. 14
({\it filled circles}) and Mar. 7 ({\it filled squares}).  Each point
represents a bin 100~\AA\ wide, with the symbol size increasing from the blue
end of the spectrum ($\lambda \approx 4000$~\AA) to the red (10100~\AA\ for
Feb. 14, and 8800~\AA\ for Mar. 7).  The best-fitting line to each epoch,
determined by a uniform-weight least-squares fit, is drawn through the points;
the point of intersection of the two lines represents a possible value of the
ISP.  When averaged with the results of another technique (see
\S~\ref{sec:removalofisp}), the adopted ISP ($q_{\rm ISP} = -0.17\%, u_{\rm
ISP} = -0.48\%$; {\it open triangle}) is derived. The general regions
containing the points representing the polarization across the \ion{Ca}{2} IR
triplet are also indicated ({\it absorption, emission}).
\label{fig:3} }
\end{figure}

One way to estimate the actual total ISP level is to rely on the data obtained
for SN~2002ap itself, along with some theoretical assumptions.
Figure~\ref{fig:3} shows the observed polarization data in the $q$-$u$ plane
for both epochs.  The majority of the data from each epoch appears to scatter
about straight, but different, lines in the $q$-$u$ plane.  Such behavior is
indicative of an axisymmetric asphericity in which the intrinsic P.A. is
wavelength independent, a trait that has been identified in previous SN
polarization studies (e.g., Howell et al. 2001; Leonard et al. 2001).
If axisymmetry is suspected, then the ISP must lie along the best-fitting line
to the points in the $q$-$u$ plane; since the dominant axis for the two epochs
prefer different P.A.s ($\sim 95^\circ {\rm\ and\ } 150^\circ$ on days 16 and
37, respectively), the intersection of the two lines provides an estimate of
the ISP.  From Figure~\ref{fig:3} we see that this occurs near $q_{\rm ISP} =
-0.19\%, \ u_{\rm ISP} = -0.53\%$, or $p_{\rm ISP} = 0.56\%, \ \theta_{\rm ISP} =
125^\circ$.  Since ISP varies with wavelength according to a ``Serkowski''
curve (Whittet et al. 1992), it has a wavelength dependence; for our purposes,
we take the derived ISP location to represent the value at $\lambda_{\rm max} =
5500$~\AA, the average wavelength of maximum ISP found from studies of Galactic
stars.

A second method to estimate the total ISP is to assume that some spectral
region is intrinsically {\it completely} unpolarized, with all of the {\it
observed} polarization therefore coming from ISP.  There is both theoretical
and empirical evidence that this may be the case in the \ion{Ca}{2} IR emission
profile on day 37.  Since directional information is lost as photons are
absorbed and reemitted in a line (H\"{o}flich et al. 1996), broad, unblended
emission features have traditionally been treated as intrinsically unpolarized
in SN studies.  In their spectropolarimetric study of an early-time,
emission-dominated spectrum of SN~1998S, Leonard et al. (2000) demonstrate that
the basic paradigm of polarized continuum light being diluted by unpolarized
emission lines is quite consistent with that data: the degree of depolarization
across the broad-line features of SN~1998S changes in direct proportion to
strength of the line at each wavelength across the profile.  In the case of the
\ion{Ca}{2} profile of SN~2002ap, however, we see that this simple conceptual
scheme fails, since the {\it entire} emission component, and even some of the
absorption component, is polarized at roughly the same level
(Fig.~\ref{fig:2}); such behavior is also seen in the \halpha\ profiles of the
Type II-P SN~1999em (Leonard et al. 2001).  One possible interpretation of this
is that not only are the emission-line photons unpolarized, but the underlying
continuum photons are as well, having been absorbed and reemitted by the
optically thick line. Such a situation is also described by Howell et
al. (2001) for an SN~Ia.  If we accept this premise, and estimate the ISP from
a 200~\AA\ wide bin centered on the emission peak at 8578~\AA, we derive
$q_{8578} = -0.12\%, \ u_{8578} = -0.35\%$.  Fitting a Serkowski law ISP curve
(Whittet et al. 1992) to this value with an assumed wavelength of maximum
polarization of $\lambda_{\rm max} = 5500$~\AA, we derive $q_{\rm ISP} =
-0.15\%, \ u_{\rm ISP} = -0.43\%$ at $\lambda = 5500$ \AA, or $p_{\rm ISP} =
0.46\%, \ \theta_{\rm ISP} = 125^\circ$.  This result is quite similar to that
derived earlier by the $q$-$u$ plane method, and is also consistent with the
value Kawabata et al. (2002) found using a similar technique ($p_{\rm ISP} =
0.64\% \pm 0.20\%, \ \theta_{\rm ISP} = 120^\circ \pm 10^\circ$).

The consistency between our two independent total ISP estimates affords some
confidence that we have determined an accurate value.  Taking the average of
the two results yields an estimate of the total ISP of $q_{\rm ISP} = -0.17\%,
\ u_{\rm ISP} = -0.48\%$, or $p_{\rm ISP} = 0.51\%$ at $\theta_{\rm ISP} =
125^\circ$.  Since the reddening from Milky Way (MW) dust is thought to
dominate the total reddening of SN~2002ap, we would expect Galactic ISP to make
up the bulk of the total ISP.  To check this assertion, we obtained
spectropolarimetry of five distant Galactic stars near to the l-o-s of the
object (the catalog of Heiles 2000 records no polarization measurements of any
Galactic star within $1^\circ$ of the l-o-s to SN~2002ap, the nominal angular
separation over which some consistency of ISP can be assumed), and list the
results in Table 2.  Although the values have some scatter, they are generally
consistent with each other (they all fall in the same quadrant of the $q$-$u$
plane) and confirm a non-negligible polarization in this area of the sky from
Galactic dust.  Calculating the mean and $1\sigma$ scatter of the individual
values, we obtain $q_{\rm MW} = -0.11 \pm 0.08\%, \ u_{\rm MW} = -0.29 \pm
0.14\%$, which yields $p_{\rm MW} = 0.32 \pm 0.13\%$ at $\theta_{\rm MW} =
125^\circ \pm 12^\circ$.  This is to be compared with the total ISP estimated
by the earlier techniques of $p_{\rm ISP} = 0.51\%, \ \theta_{\rm ISP} =
125^\circ$.  The general agreement of all of these values affirms our
contention that Galactic ISP makes up the bulk of the total ISP, and also
strengthens our belief that the total ISP estimated by the earlier techniques
must be nearly correct.  We therefore adopt $p_{\rm ISP} = 0.51\%$ at
$\theta_{\rm ISP} = 125^\circ$ for $\lambda_{\rm max} = 5500$~\AA, and proceed
to remove this ISP from our data.\footnote{The shape of some of the Galactic
stars' polarization spectra suggests that $\lambda_{\rm max}$ may be somewhat 
less than the nominal value of $\lambda_{\rm max} = 5500$~\AA.  Given the inherent
difficulty of accurately measuring such low polarization values (see, e.g.,
Leonard et al. 2001), however, we choose to use the nominal value of
$\lambda_{\rm max} = 5500$ \AA\ to correct our data.}


\begin{figure}
\ssp
\begin{center}
\rotatebox{0}{
\scalebox{1.0}{
\plotone{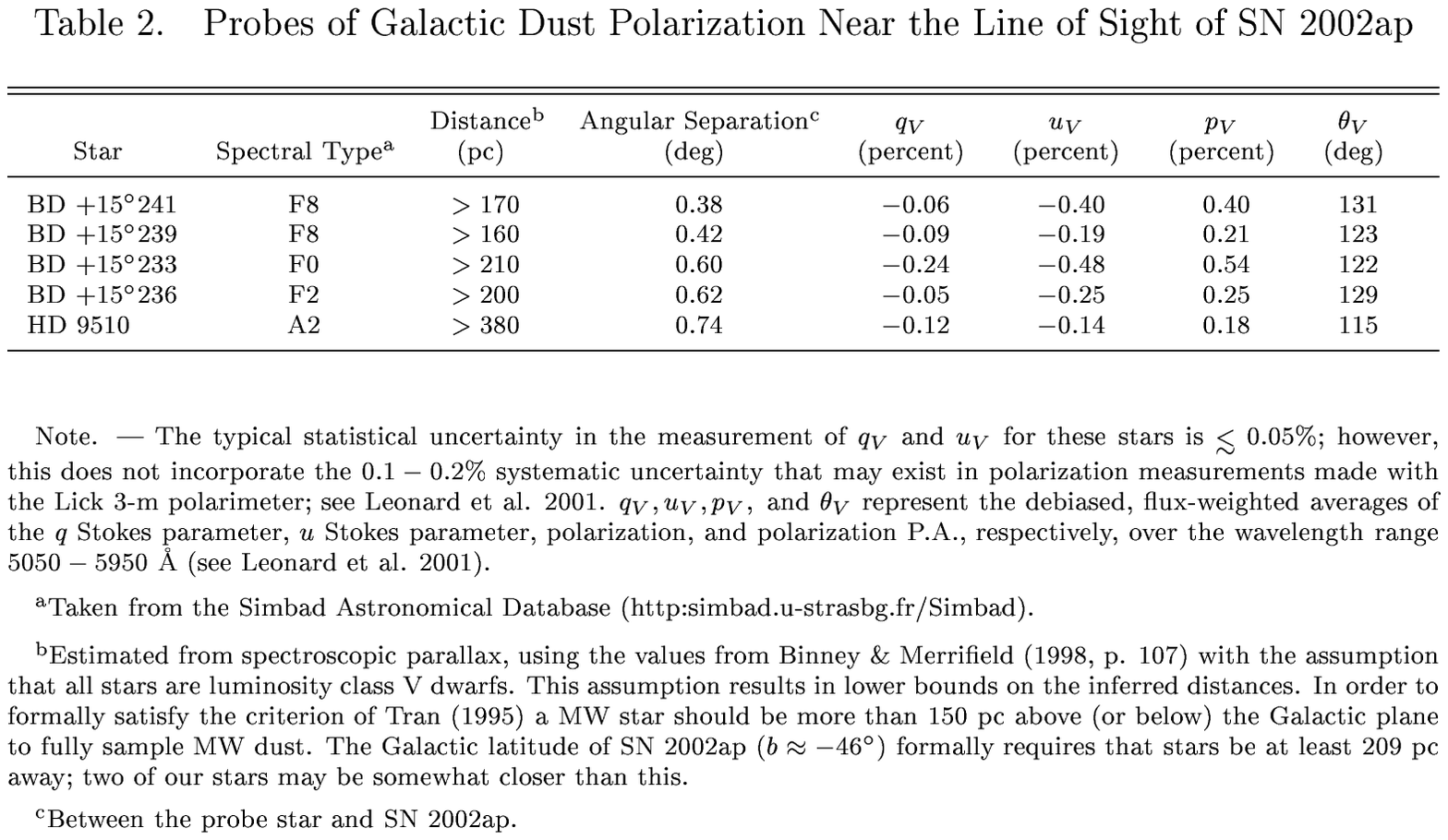}
}
}
\end{center}
\end{figure}

\subsection{Intrinsic Polarization}
\label{sec:intrinsicpolarization}

\begin{figure}
\ssp
\begin{center}
\rotatebox{90}{
\scalebox{0.7}{
\plotone{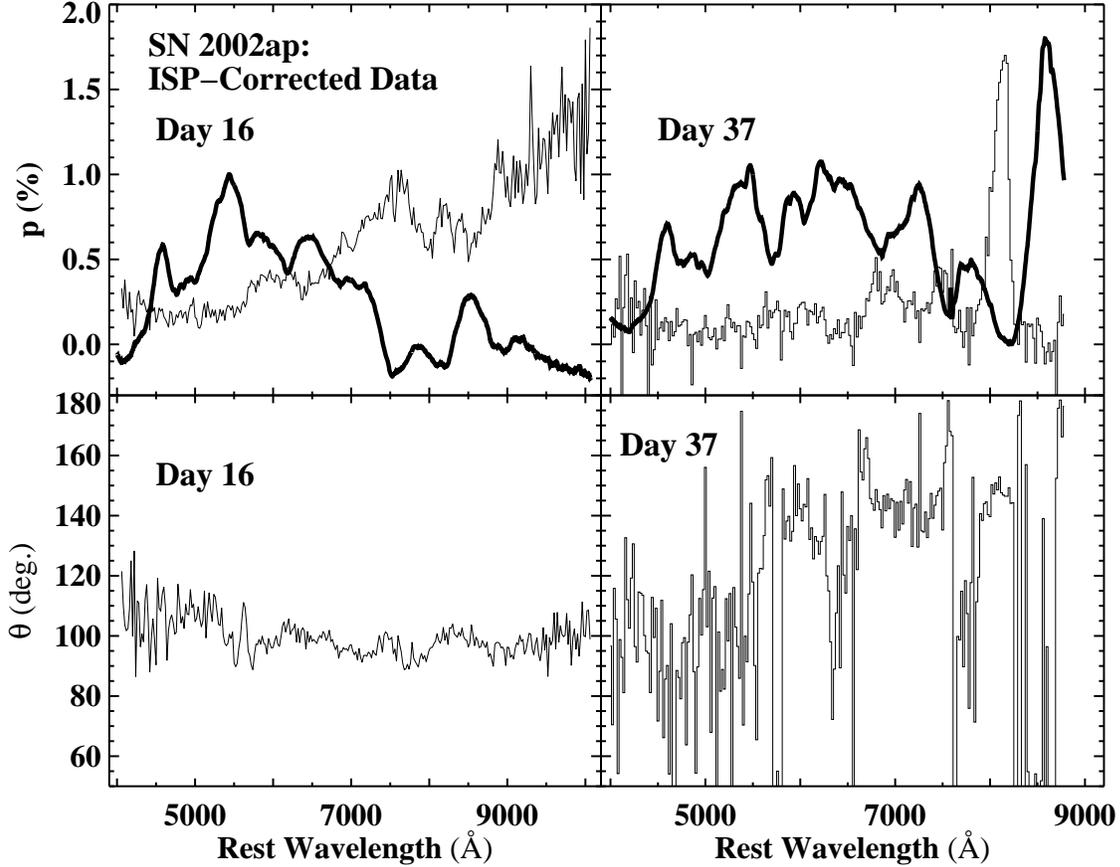}
}
}
\end{center}
\caption{Polarization level ({\it top, thin lines}) and P.A. ({\it bottom})
intrinsic to SN~2002ap after removal of an ISP characterized by $p_{\rm ISP} =
0.51\%$ at $\theta_{\rm ISP} = 125^\circ$ for days 16 ({\it left panels}) and
37 ({\it right panels}) after explosion.  For comparison of features,
arbitrarily scaled total flux spectra ({\it top panels, thick lines}) are
overplotted.  Since the very low inferred polarization for much of the observed
spectral range in the day 37 epoch renders the P.A. quite indeterminate and an
RSP difficult to define, we display the ``optimal'' polarization, defined as $
p_{trad} - [(\sigma^2(p_{trad}))/p_{trad}]$, where $p_{trad} = \sqrt{q^2 +
u^2}$ (see discussion in Leonard et al. 2001, and references therein) for the
day 37 data.
\label{fig:4} }
\end{figure}

The resulting ISP-corrected, intrinsic SN polarization and P.A. are displayed
in Figure~\ref{fig:4}.  The outstanding feature of the intrinsic polarization
on day 16 is its strong increase with wavelength, from $p \approx 0.2\%$ near
4000~\AA\ to $p \approx 1.3\%$ at the cutoff near 10,000~\AA.  Such a
precipitous rise in polarization at optical and near-infrared wavelengths is
completely unlike the Serkowski law dependence of ISP, and must be intrinsic to
the SN.  A similar redward rise in intrinsic polarization was seen in
spectropolarimetry of the Type Ia SN~1999by, where it was attributed to the
decreasing importance of line opacities, and the increased influence of
continuum electron scattering at longer wavelengths (Howell et al. 2001).  In
particular, numerous lines, largely due to iron-group elements, dominate the
spectrum at $\lambda \lesssim 6150$~\AA\ and significantly depolarize the
light, whereas the red end of the spectrum suffers less severe line-blending,
allowing the intrinsically polarized continuum light to dominate.  We suspect
that a similar effect may be operating here, and that the high polarization ($p
\approx 1.3\%$) measured at the red end of the spectrum in fact represents a
lower limit on the true level of continuum polarization undiluted by
unpolarized line photons at this epoch.


\begin{figure}
\ssp
\begin{center}
\rotatebox{0}{
\scalebox{0.7}{
\plotone{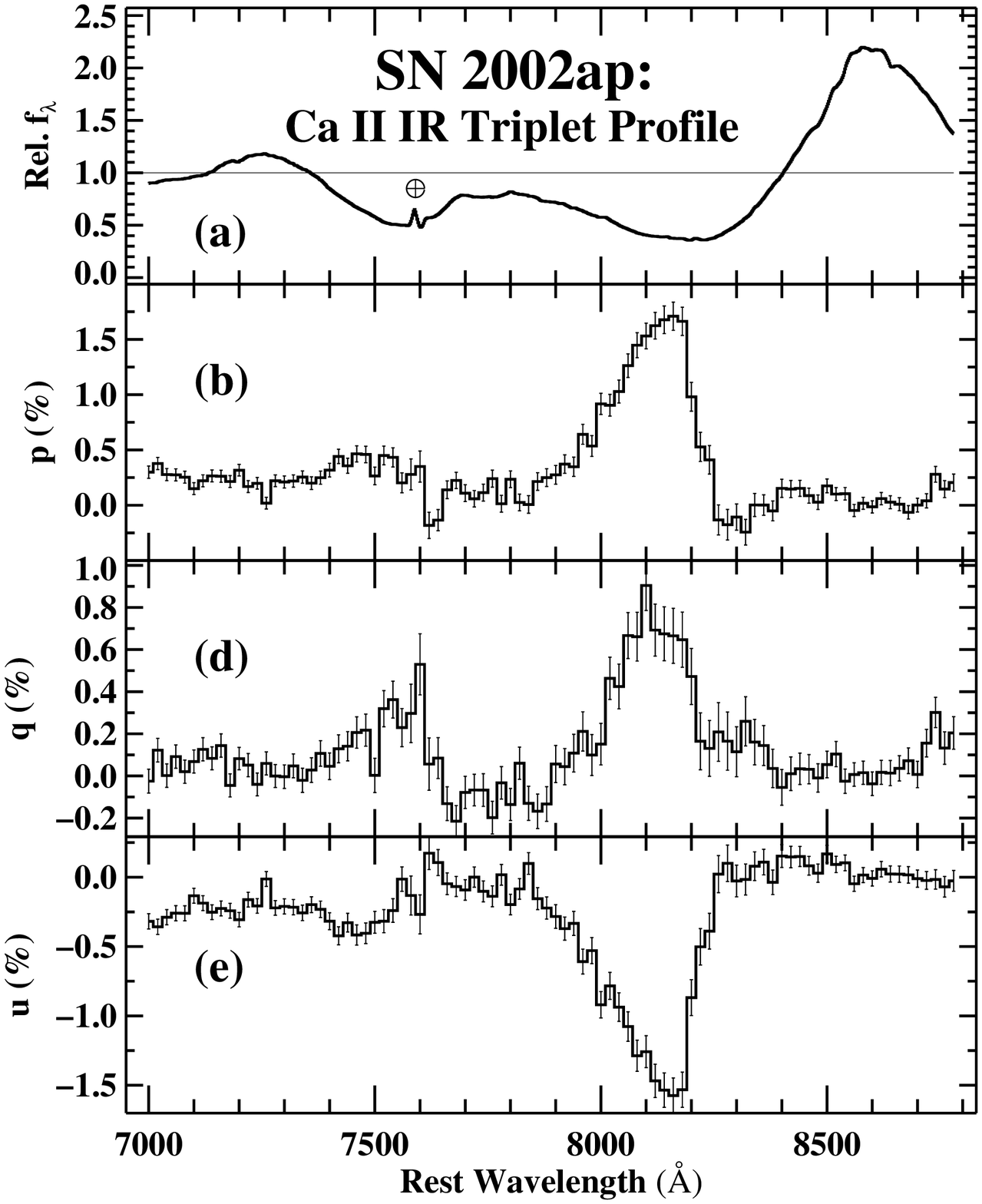}
}
}
\end{center}
\caption{ISP-corrected spectropolarimetry of the region near the \ion{Ca}{2} IR
triplet for SN~2002ap on 2002 March 7, roughly 37 days after explosion.  Error
bars are $1\sigma$ statistical for 20~\AA\ bin$^{-1}$.  ({\it a}) Total flux,
normalized by a spline fit to the continuum, displayed at 2~\AA\ bin$^{-1}$ for
better resolution.  ({\it b}) Intrinsic degree of polarization.  ({\it c, d})
Normalized $q$ and $u$ Stokes parameters.  The spurious flux change near
7600~\AA\ is due to the difficulty in accurately removing the A-band telluric
feature from the high air-mass observations.
\label{fig:5} }
\end{figure}

A detail of the region near the strong polarization increase associated with
 the absorption trough of the \ion{Ca}{2} IR triplet on day 37 is shown in
 Figure~\ref{fig:5}.  In the absorption trough there is an apparent change in
 observed polarization of $\Delta p_{tot} = 1.8\%$, where

\begin{equation}
\Delta p_{\rm tot} \equiv \sqrt{\Delta q^2 + \Delta u^2},
\label{eqn:1}
\end{equation}

\noindent and $\Delta q\ {\rm and\ } \Delta u$ are the individual changes in
the normalized Stokes parameters compared with the nearby continuum value (note
that $\Delta p_{\rm tot}$ is an {\it ISP-independent} quantity; see Leonard \&
Filippenko 2001).  It is possible to interpret this sharp polarization increase
within the basic model first proposed by McCall (1984) and later extended by
Jeffery (1991), that polarization peaks are naturally associated with
absorption minima due to selective blocking of forward-scattered (and hence
less polarized) light in P-Cygni troughs.  With this simple model in mind,
Leonard \& Filippenko (2001) showed that a lower bound could be placed on the
true intrinsic continuum polarization level of

\begin{equation}
p_{\rm cont} \geq \frac{\Delta p_{\rm tot}} {(I_{\rm cont}/I_{\rm trough}) - 1},
\label{eqn:2}
\end{equation}

\noindent where {$\Delta p_{tot}$} is defined by Equation~\ref{eqn:1}, $I_{\rm
cont}$ equals the interpolated value of the continuum flux at the location of
the line trough, and $I_{\rm trough}$ is the total flux at the line's flux
minimum.  The main difficulty in the application of this to SN~2002ap is
interpolating the continuum flux across the \ion{Ca}{2} profile; our adopted
estimate of the continuum level is indicated in Figure~\ref{fig:5}a.  With
$\Delta p_{\rm tot} \approx 1.8\%$ and $I_{\rm cont}/I_{\rm trough} \approx
2.8$, we find that $p_{\rm cont} > 1.0\%$ for this epoch.

We therefore see evidence for significant intrinsic continuum polarization on
both day 16 and day 37 after explosion.  When modeled in terms of the oblate,
electron-scattering atmospheres of H\"{o}flich (1991), the inferred continuum
polarization demands an asphericity of at least $20\%$ on day 16 and $17\%$ on
day 37; the actual value naturally depends on the (unknown) viewing
orientation.  It is also possible that ionization asymmetry or clumping in the
ejecta may contribute to the polarization, which would reduce the inferred bulk
asymmetry (see, e.g., H\"{o}flich, Khokhlov, \& Wang 2001).  Detailed
theoretical modeling of both the total flux and polarization data may set
constraints on the viewing orientation (e.g., H\"{o}flich, Wheeler, \& Wang
1999), and also serve to check the specific assumptions that underlie our
estimates of the intrinsic continuum polarization.

\subsection{Nebular Flux Spectrum}
\label{sec:nebularfluxspectrum}

\begin{figure}
\ssp
\begin{center}
\rotatebox{90}{
\scalebox{0.7}{
\plotone{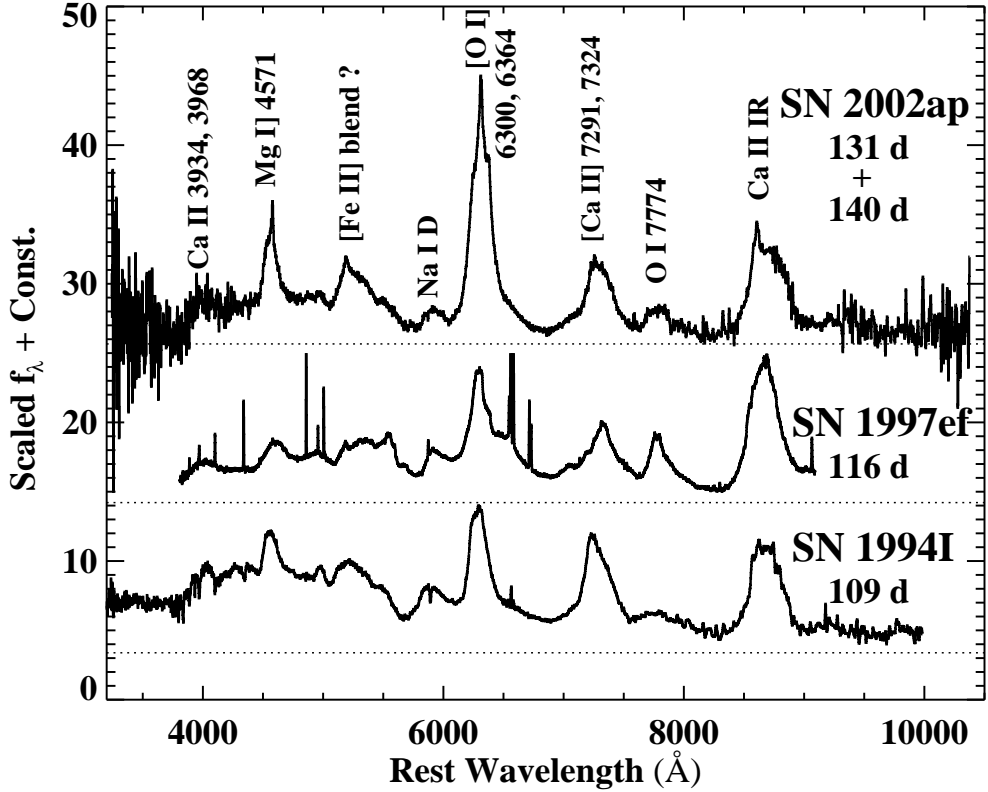}
}
}
\end{center}
\caption{SN~2002ap ({\it top}) in the early nebular phase, compared with the
peculiar Type Ic ``hypernova'' SN~1997ef ({\it middle}) and the ``ordinary''
Type Ic SN~1994I ({\it bottom}) at similar epochs with estimated day since
explosion indicated.  The spectra are scaled so that the strengths of the
\ion{Ca}{2} IR lines are approximately the same.  Prominent emission lines are
identified, and the zero-flux levels ({\it dotted lines}) are indicated.  The
[\ion{Fe}{2}] blend near $5200$ \AA\ may contain significant [\ion{O}{1}] and
[\ion{Mg}{1}] emission at this early nebular epoch (see discussion by Maeda et
al. 2002).  Some of the narrow emission lines resulting from a superposed
\ion{H}{2} region have been clipped in the spectrum of SN~1997ef.
\label{fig:6} }
\end{figure}

Figure~\ref{fig:6} shows the total flux spectrum of SN~2002ap obtained after
the SN reappeared following solar conjunction; we have combined the
observations taken 131 and 140 days after the explosion in order to improve
the signal-to-noise ratio.  The spectrum is characterized by strong emission
profiles superimposed on a very weak continuum, implying that the SN has
entered the nebular phase (Maeda et al. 2002; Sollerman et al. 2000).  The
greater strength of the lines relative to the continuum level in the spectrum
of SN~2002ap compared with other SNe Ic at similar, although slightly earlier
epochs (Fig.~6), is consistent with the rapid early spectral evolution reported
by Mazzali et al. (2002), who found that SN~2002ap evolved at about twice the
rate of SN~1997ef.

Early-time spectra of SN~2002ap showed very broad spectral features, indicating
high velocities in the outer ejecta ($\sim 30,000$ \kms; Mazzali et al. 2002).
Given the extreme breadth of the early-time spectral features, one might expect
the nebular-phase broad-line widths (taken as the FWHM of a Gaussian fit to the
continuum-subtracted line profile; see Matheson et al. 2001) for SN~2002ap to
be greater than those measured in other nebular SNe~Ic.  However, for
[\ion{O}{1}] $\lambda\lambda$6300, 6364 we measure 8600~\kms, consistent with
the average value ($7600 \pm 2100$ \kms) of 9 SNe Ic reported by Matheson et
al. (2001), and for [\ion{Ca}{2}] $\lambda\lambda$7291, 7324, we measure
9800~\kms, also consistent with the Matheson et al. (2001) value ($8700 \pm
2700$ \kms).  At this stage, there is evidently not significantly more line
emission coming from high-velocity gas in this ``hypernova'' than is typical
for ``ordinary'' SNe~Ic.  Similarly normal expansion velocities were also found
at late times for the hypernova SN~1998bw (Patat et al. 2001).  In addiion, the
identification of spectral features extending out to nearly $\sim 30,000$ \kms\
in very early-time spectra of the Type II-P SN~1999gi (Leonard et al. 2002b)
suggests that the presence of high-velocity material may be common among
core-collapse SNe, but is only easily identified if very early-time spectra are
obtained.


\begin{figure}
\ssp
\begin{center}
\rotatebox{0}{
\scalebox{0.7}{
\plotone{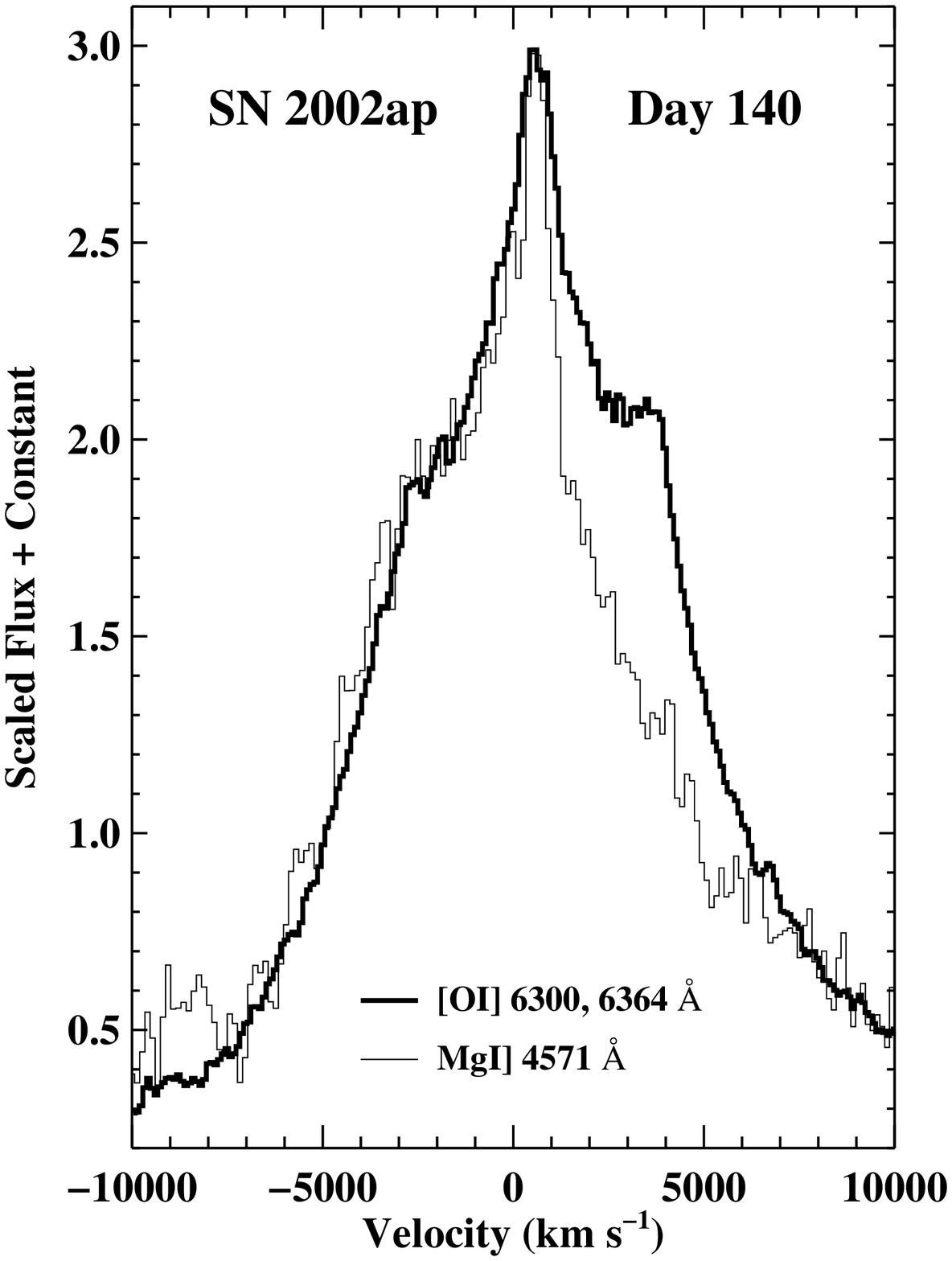}
}
}
\end{center}
\caption{The [\ion{O}{1}] $\lambda\lambda 6300, 6364$ ({\it thick line}) and
\ion{Mg}{1}] $\lambda 4571$ ({\it thin line}) profiles of SN~2002ap, 140
days after explosion.  In addition to the broad lines, there are narrow
emission spikes in both profiles, redshifted by $\sim 580$~\kms\ with respect
to the systemic velocity of M74.  
\label{fig:7} }
\end{figure}

Figure~\ref{fig:7} shows the [\ion{O}{1}] $\lambda\lambda$6300, 6364 and
\ion{Mg}{1}] $\lambda$4571 line profiles for day 140 (shown instead of the day
131 spectrum due to its superior resolution).  A striking feature of both
profiles is the very sharp peaks to the flux distribution, which may indicate
the presence of a narrow-line component, redshifted by about 580 \kms\ with
respect to the systemic velocity of the host galaxy M74.  Given the complex
nature of the profiles, it is not certain whether these narrow features are
resolved in our spectrum, which has a resolution of $\Delta v \approx 400$
\kms\ at 6300~\AA.  Patat et al. (2001) observed redshifted emission peaks in
spectra of SN~1998bw at a similar phase.  This emission could come from an
excess of gas existing in lower-velocity regions close to the center of the SN
nebula.  It is also possible that it is coming from material previously lost by
the progenitor star and now existing (in clumps?) in the circumstellar
environment of the SN (e.g., SN~1999cq; Matheson et al. 2000a).  Continued
monitoring at sufficient spectral resolution is urged in order to trace the
development of these narrow features.

Given the spectropolarimetric evidence for asphericity during the photospheric
phase (\S~\ref{sec:intrinsicpolarization}), it is worthwhile to investigate
whether the nebular-phase line profiles show any evidence for asymmetry in the
line-emitting regions at these later epochs.  As discussed in
\S~\ref{sec:introduction}, the main spectral diagnostic of asymmetry was found
by Maeda et al. (2002) to be a large ratio ($\gtrsim 1.5$) between the width of
the [\ion{Fe}{2}] blend near 5200~\AA\ and that observed in [\ion{O}{1}]
$\lambda\lambda$6300, 6364 (see Fig.~\ref{fig:6} for the identification of
these features).  One difficulty with this diagnostic, however, is the complex
nature of the [\ion{Fe}{2}] blend's profile, which might include contributions
from other lines (e.g., [\ion{O}{1}] $\lambda 5577$; [\ion{Mg}{1}] $\lambda
5173$) that can substantially widen the observed profile, especially during the
early nebular phase.  It is also difficult to estimate the ``continuum''
underlying this feature, since it evidently drops significantly across the
profile.  In their study, Maeda et al. (2002) analyzed the feature in a nebular
spectrum of SN~1998bw taken 201 days after $B$-band maximum, in which they felt
confident that no extraneous lines contributed to the profile.  Given the rapid
spectral development of SN~2002ap, it is possible that this region in our
spectrum, from $\sim130$ days after $B$ maximum, is similarly uncontaminated.
Detailed spectral models are needed to confirm this assumption, but they are
beyond the scope of this paper.  If we assume that our feature is the same as
that observed in SN~1998bw at the later epoch, then we derive a velocity ratio
of $\sim$ 2.8:1 for the features.  This value is similar to that found for
SN~1998bw, which would therefore imply an aspherical explosion for SN~2002ap in
the same way that it was inferred for SN~1998bw.  However, the relatively early
phase of our spectrum, the difficulty inherent in measuring the width of the
[\ion{Fe}{2}] feature, and the potential for significant line blending make us
view this preliminary result with some degree of skepticism.

\section{Discussion and Conclusions}
\label{sec:conclusions}

There is strong evidence from spectropolarimetry and tentative evidence from
nebular-phase line profiles that SN~2002ap is not spherical.  It is useful to
compare the spectropolarimetry of SN~2002ap with the small but growing number
of core-collapse events for which polarimetry measurements exist.  The lower
bounds on the continuum polarization level of SN~2002ap ($1.3\%$ and $1.0\%$ on
days 16 and 37, respectively) are larger than those typically derived for SNe
II-P (Wheeler 2000), but not nearly as great as that inferred for the SN~Ic
1997X (Wang et al. 2001).\footnote{The presentation of data for SN~1997X by
Wang et al. (2001) shows broadband averages of the polarization data.  It is
therefore unknown what fraction of the claimed polarization ($p > 4\%$) might
be due to line-feature modulations similar to what is seen in the \ion{Ca}{2}
IR triplet of SN 2002ap during the second epoch, and not to intrinsically
polarized continuum light.}  A more direct comparison to a similar event can be
made with the spectropolarimetry presented by Patat et al. (2001) for the
hypernova SN~1998bw.  Patat et al. (2001) present spectropolarimetry from two
epochs, $-7$ and +10 days after $B$-band maximum (our observations of SN~2002ap
occurred on days 7 and 28 relative to the date of $B$ maximum derived by
Gal-Yam et al. 2002) with coverage from 4000~\AA\ to 6900~\AA.  Although the
ISP is unknown for this event,\footnote{The Galactic ISP derived by adopting
the measured linear polarization of a nearby star, HD 184100 (Kay et al. 1998),
is probably not appropriate since this star has a spectroscopic parallax of
only $\sim 14$ pc, which is insufficient to fully sample the dust in the
Galactic plane.} we note that the overall level of the observed polarization
($p \approx 0.5 - 0.7\%$), and the marked increase in the polarization level
from blue to red wavelengths (particularly in the earliest epoch), are quite
similar to the behavior of SN~2002ap.

In addition to the high inferred continuum polarization, another interesting
finding for SN~2002ap is that while the data from both epochs suggest a
substantial degree of axisymmetry (Fig.~\ref{fig:3}), the dominant axes have
quite different polarization angles, changing from $\sim 95^\circ$ on day 16 to
$\sim 150^\circ$ on day 37 (\S~\ref{sec:removalofisp}).  This behavior is
different from that observed in other core-collapse events, most notably those
resulting from progenitors with massive envelopes intact at the time of
explosion.  For SN~1999em, a classic Type II-P event, Leonard et al. (2001)
show that a common, wavelength-independent P.A. exists for all observational
epochs, which includes days 12, 45, 54, 164, and 168 after explosion.  This
suggests a more complex morphology for the thinning ejecta of SN~2002ap.

As mentioned in \S~\ref{sec:introduction}, a potentially relevant result of
recent simulations involving explosive nucleosynthesis in aspherical,
jet-induced SN explosions is that intermediate-mass and heavy elements (e.g.,
Fe) as well as freshly synthesized radioactive material such as $^{56}$Ni are
ejected (at high velocity) primarily along the poles whereas elements
synthesized in the progenitor (e.g., He, C, Ca, O) are preferentially located
at lower velocities near the equatorial plane in the expanding ejecta (Maeda et
al. 2002; Khokhlov \& H\"{o}flich 2001; H\"{o}flich et al. 2001).  Although the
P.A. in the ISP-corrected data from day 37 (bottom, right panel of
Fig.~\ref{fig:4}) is difficult to define in some regions due to the low
polarization level, there is evidently a change near $\lambda \approx 6000$
\AA.  Below this wavelength, $\theta \approx 100^\circ$, a value similar to the
one that characterizes the day 16 data.  Above $\sim 6000$ \AA, $\theta \approx
150^\circ$ tends to dominate; this is especially easy to see (and believe, due
to the high polarization) in the \ion{Ca}{2} IR absorption trough (see also
Fig.~\ref{fig:3}).  Might this change in P.A. be due to the different spatial
distributions of the iron-group elements, which dominate at blue wavelengths,
from the elements synthesized by the progenitors, such as Ca and O, that are
responsible for much of the opacity at red wavelengths (see Fig.~\ref{fig:2}a)?
Along these same lines, an explanation for the nearly constant P.A. on day 16
at all wavelengths, and its similarity to the P.A. of day 37 at blue
wavelengths, could be due to the photosphere at early times being located in
the highest-velocity (i.e., polar) material. By day 37, perhaps the SN has
cooled sufficiently to expose the inner, slower-moving equatorial regions, and
we are witnessing the competition between these two different distributions of
material.

To be sure, a small change in the ISP estimate, or even the wavelength of
maximum polarization of the ISP, could alter the inferred P.A. of the continuum
in the second epoch considerably.   Further interpretation of the polarization
change near $6000$ \AA\ in the second epoch is probably not warranted.  The
$\sim 55^\circ$ P.A. difference between the the day 16 data and the region
associated with the \ion{Ca}{2} IR trough on day 37, however, remains
regardless of the ISP choice, and may well indicate a true difference in the
distribution of Ca relative to iron-group elements.


\begin{figure}
\ssp
\begin{center}
\rotatebox{0}{
\scalebox{0.7}{
\plotone{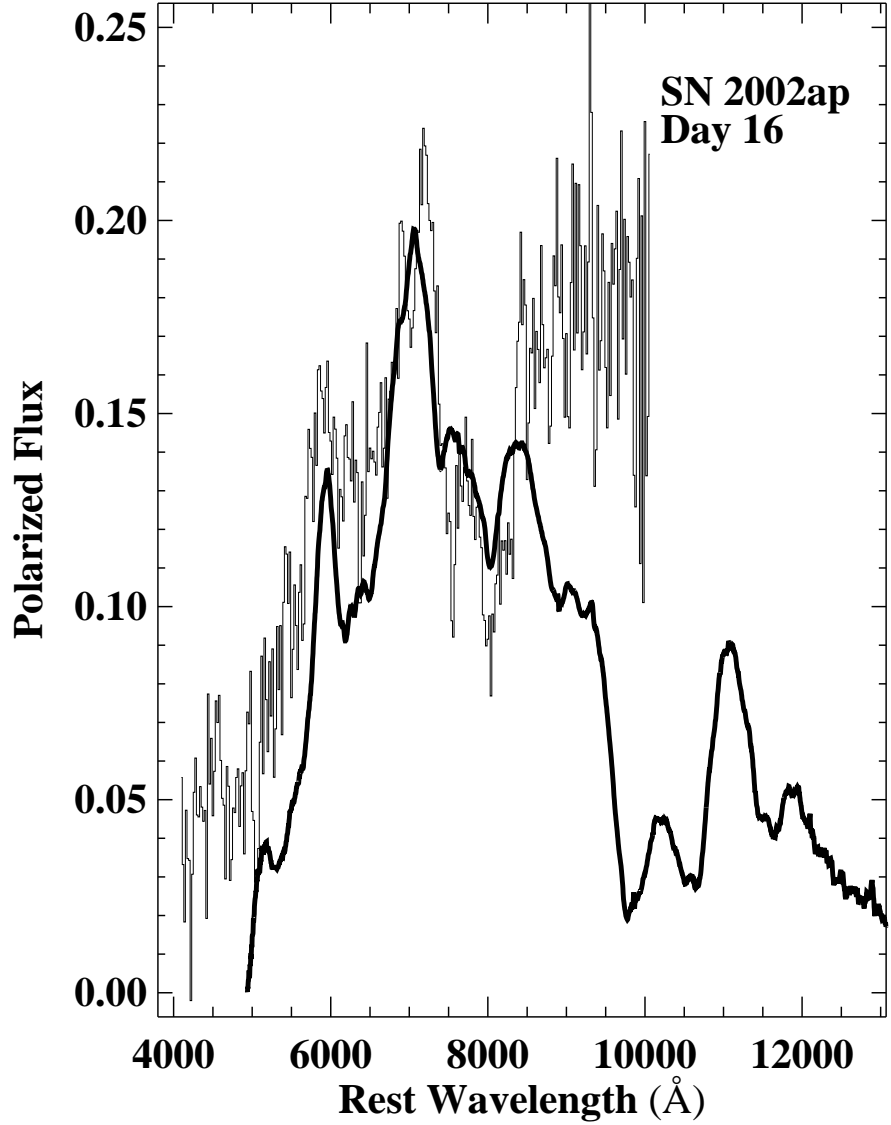}
}
}
\end{center}
\caption{The polarized flux of SN~2002ap on 2002 Feb. 14 (16 days after
explosion; {\it thin line}) compared with the total flux spectrum from
the same date redshifted by 0.23$c$ ({\it thick line}), arbitrarily scaled
and offset for comparison of features. 
\label{fig:8} }
\end{figure}

Just prior to completion of this research, a paper describing similar data
appeared on astro-ph by Kawabata et al. (2002).  Although the observational
epochs are slightly different (Kawabata et al. present data from 2002 February
9.2--11.3 and March 8.2--10.2), the general results are similar: an observed
polarization level in the February epochs of $p \approx 0.6\%$ that may rise
somewhat toward the red (the increase is not as obvious as it is in our data
since their data does not extend beyond 8300~\AA) and a dramatic polarization
increase in the \ion{Ca}{2} IR triplet's absorption trough followed by
depolarization in the emission profile in the March data.  One interesting
speculation made by Kawabata et al. (2002) is that much of the polarized
continuum at early times may result from scattering off of electrons in a jet,
or bipolar jets, of material emitted from the SN during the explosion.
Supporting this hypothesis is the general similarity demonstrated between the
intrinsic polarized flux (i.e., $p_{\rm intrinsic} \times f$) and the total
flux spectrum redshifted by $0.23c$, the presumed speed of the jet.  We also
find qualitative agreement between the intrinsic polarized flux and the total
flux redshifted by $z = 0.23c$ during our February epoch (Fig.~\ref{fig:8}),
although we have not investigated how significant the correlation is.  We do
note, though, that the speculation by Kawabata et al. (2002) is consistent with
the scenario proposed above that the polarization P.A. on day 16 is dominated by
the distribution of the polar jet material.  Countering the jet hypothesis,
however, is the recent assertion by Berger et al. (2002) that the extreme
faintness of SN~2002ap at radio wavelengths is inconsistent with the presence
of a jet, regardless of the viewing geometry.  In any event, additional
spectropolarimetry of young SNe Ic is certainly warranted in order to further
test the jet model.

The total flux spectra obtained in the early-nebular phase confirm that
SN~2002ap continues the rapid spectral evolution previously witnessed during
the photospheric phase by Mazzali et al. (2002).  It is characterized by a very
weak continuum and broad emission lines, which demonstrate that it is already
well along the transition to the fully nebular phase.  The lines have widths
similar to those observed in ``normal'' nebular SNe Ic.  There are, however,
unusual narrow lines superimposed on some of the broad-line profiles, including
especially those of [\ion{O}{1}] $\lambda\lambda$6300, 6364 and \ion{Mg}{1}]
$\lambda$4571.  It is possible that the ratio of the width of the [\ion{Fe}{2}]
blend near 5200~\AA\ compared with that of [\ion{O}{1}] $\lambda\lambda$6300,
6364 is indicative of asymmetry in the line-emitting region, but additional,
later-time spectra are needed to confirm this result.

\acknowledgments

Some of the data presented herein were obtained at the W. M. Keck Observatory,
which is operated as a scientific partnership among the California Institute of
Technology, the University of California, and the National Aeronautics and
Space Administration.  The Observatory was made possible by the generous
financial support of the W. M. Keck Foundation.  This research has made use of
the NASA/IPAC Extragalactic Database (NED), which is operated by the Jet
Propulsion Laboratory, California Institute of Technology, under contract with
NASA, as well as the SIMBAD database, operated at CDS, Strasbourg, France.
This research was supported in part by NASA through the American Astronomical
Society's Small Research Grant Program.  Our work was also funded by NASA
grants GO-8648, GO-9114, and GO-9155 from the Space Telescope Science
Institute, which is operated by AURA, Inc., under NASA contract NAS 5-26555.
Additional funding was provided to A.V.F. by NSF grant AST-9987438. We thank
A. J. Barth and E. C. Moran for help with some of the observations, W. Li for
useful discussions, and an anonymous referee for helpful suggestions that
resulted in an improved manuscript. The assistance of the staffs at Lick and
Keck Observatories is greatly appreciated.

\clearpage

\appendix
\section{APPENDIX}
\subsection {Details of the Observations}

The observational approach that we have taken at Keck in the past for our
program to study SN geometry has been to obtain spectropolarimetry with a 600
groove mm$^{-1}$ grating blazed at 5000~\AA, which yields a spectral range of
about 4350--6850 \AA\ in the red arm of LRIS (Leonard et al. 2000; Leonard et
al. 2001; Leonard \& Filippenko 2001).  The observing procedure of SN~2002ap on
February 14 and March 7 differed in two ways from our traditional approach.
First, in an effort to take advantage of the extended coverage possible with
the recently commissioned blue arm of the LRIS double spectrograph (LRIS-B; an
overview of the LRIS-B instrument is available at
\url{http://alamoana.keck.hawaii.edu/inst/lris/lrisb.html}), on February 14 we
used the D680 dichroic filter to split the spectrum at 6800~\AA\ and send the
blue and red light to the respective arms of LRIS (hereafter referred to as the
``UV'' and ``IR'' settings, respectively).  To our knowledge, these are the
first spectropolarimetric observations to be published using LRIS-B.  Second,
on both nights we observed SN~2002ap using the red side alone (i.e., no
dichroic) with a 300 groove mm$^{-1}$ grating blazed at 5000~\AA, which yielded
coverage in the range 3930--8830~\AA\ (hereafter referred to as the ``OPT'' setting).
The lack of an order-blocking filter in this setting makes second-order light
contamination beyond $\sim 6400$~\AA\ a concern.  In this Appendix, we discuss
the performance of LRIS-B for spectropolarimetry and the impact of second-order
light on the data acquired with the OPT setting.

For the UV/IR setup, a grism with 300 groove mm$^{-1}$ blazed at $5000$~\AA\
dispersed the light on the blue side, and a grating with 400 groove mm$^{-1}$
blazed at $8500$~\AA\ dispersed the light on the red side; an OG570
order-blocking filter was also inserted on the red side.  In principle, such a
setup yields coverage down to the atmospheric cutoff at $\sim$3200~\AA.
However, we found the calibration of the spectropolarimetry to be problematic
at blue wavelengths.  At these wavelengths the poor efficiency of the blue CCD 
chip and grism (see
\url{http://alamoana.keck.hawaii.edu/inst/lris/ccdblue\_qe.html} and\\
\url{http://alamoana.keck.hawaii.edu/inst/lris/grism\_list.html}), along with
the inefficiency of the ``HN-22 UV/optical Polaroid'' polarizing filter, which
is used to make the wavelength-dependent position angle correction curve (see
Miller et al. 1988; Ogle et al. 1999), probably
contribute to the difficulty we encountered.  When we attempted to calibrate
the polarization standard BD $+59^\circ389$ (Schmidt et al. 1992), strong,
artificial P.A. rotations were evident below $\sim 3800$~\AA, indicating
that the derived P.A. correction curve was not appropriate at these
wavelengths.  The data were also extremely noisy, especially for red objects.
Fortunately, a new polarizer for the blue side that should allow calibration of
polarization further into the blue has recently been installed and is
currently being tested.  In the end, we concluded that reliable
spectropolarimetry on the blue side was possible for SN~2002ap down to
3800~\AA.  The total flux spectrum was calibratable down to 3400~\AA.

An additional difficulty with the UV setting is the known problem of ``ghost''
images of second-order dispersed red-side light appearing on the blue side.
Although the intensity of the ``ghost'' images is reported to be no more than
$0.5\%$ of the red-side intensity, such contamination, if variable between
observations, could corrupt spectropolarimetry results in the UV setting.
These uncertainties, as well as some previous problems that we have had
calibrating spectropolarimetry data obtained with a dichroic (see, e.g.,
Leonard et al. 2001), led us to approach the LRIS-B spectropolarimetry with
some caution.  The OPT setting fortunately provides a check on the UV data from
this first epoch.  In a similar way, the IR setting provides a check on
possible second-order light contamination effects at red wavelengths in the OPT
setting.

On the red side, the ``IR800 IR'' polarizer, which is used to make the
wavelength-dependent position angle correction curve, does not produce $100$\%
polarized light at wavelengths beyond $\sim 8800$~\AA.  However, unlike the
case on the blue side, consistent results (i.e., no unusual P.A. or
polarization level changes) were obtained for all objects and standards out to
$1\ \mu$m, indicating that the derived P.A. correction curve is accurate.
This builds confidence in the calibration of the spectropolarimetry of
SN~2002ap at red wavelengths as recorded by the IR setting.

\begin{figure}
\ssp
\begin{center}
\rotatebox{90}{
\scalebox{0.7}{
\plotone{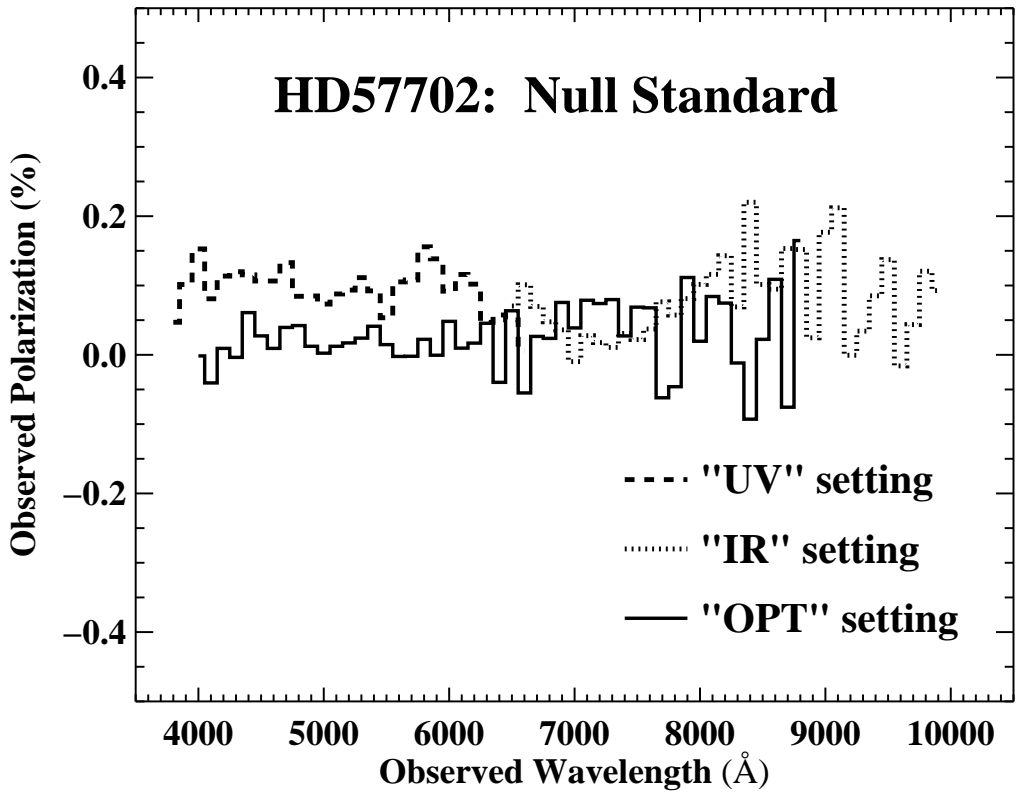}
}
}
\end{center}
\caption{Polarization of the null standard HD 57702 (Mathewson \& Ford 1970) as
recorded by the three settings described in the text, shown at 100 \AA\
bin$^{-1}$.  Polarization degree was determined by the ``optimal'' polarization
estimator (see caption to Fig.~\ref{fig:4}, and Leonard et al. 2001).
\label{fig:9} }
\end{figure}

To check for instrumental polarization, the null standard HD 57702 (Mathewson
\& Ford 1970) was observed on February 14 with all three settings
(Fig.~\ref{fig:9}).  The polarization measured with both the OPT and IR
settings are in good agreement and effectively null.  Also impressive is the
apparently good calibration of the IR polarization through the region affected
by the dichroic cutoff.  Similarly promising results were also obtained for the
polarization standard BD $+59^\circ389$ (Schmidt et al. 1992) and
other science objects observed on this night.  Although the signal-to-noise
ratio decreases sharply for both UV and IR near 6800~\AA, it may be possible to
trust the polarization data in this region with no gap in coverage.


The comparison between the UV and OPT settings for HD 57702, however, reveals
that a small ($\lesssim 0.1\%$) amount of instrumental polarization likely
exists in the UV setting.  This is also supported by similar discrepancies
between the UV and OPT settings in other objects' data on this night.  For HD
57702, Mathewson \& Ford (1970) list $p_V = 0.04 \pm 0.07\%$.  From the OPT
setting data we measure $p_V = 0.02 \pm 0.02\%$ (statistical error only), while
the UV setting yields $p_V = 0.10 \pm 0.01\%$.  Correcting for instrumental
polarization at the $0.1\%$ level is usually of dubious merit, since the
intrinsic polarization of null standards is generally of this order.
Nonetheless, since the OPT setting value is closer to the cataloged value, and
we have never found significant instrumental polarization to be evident on the
red side of LRIS, we attempted to correct for this small degree of instrumental
polarization in the UV data.  To do this, we first derived the offsets in the
raw Stokes parameters between the UV and OPT measurements of HD 57702, and then
used these offsets to correct the other objects' UV data (see Leonard et
al. 2001).  We found that this correction consistently improved the agreement
between the two settings, although in some cases the agreement between the two
settings was still not perfect.  Since improvement was seen with the
correction, though, we adopted it and applied it to the UV data obtained for
SN~2002ap.  The ultimate cause of this small amount of instrumental
polarization in our UV setting is unknown.

Although spectropolarimetry is routinely obtained at Keck using settings nearly
identical to our OPT setting (e.g., Brotherton et al. 2000; Ogle et al. 1999),
this was the first time that we personally had obtained data significantly
beyond 7000~\AA\ without the use of an order-blocking filter.  This lead to
some concern over the possibility of second-order light contaminating the data
beyond $\sim 6400$~\AA.  The degree to which second-order light affects
red-side data has been investigated in detail by Putney \& Cohen (1996) for the
300 groove mm$^{-1}$ grating in the red camera, and they show that the
percentage of blue light appearing as second-order red light in the range
6400--9000~\AA\ is at most $1.8\%$ of the incident blue flux.  Furthermore,
from observations of polarized stars made with and without an order-blocking
filter in place, Putney \& Cohen (1996) conclude that second-order light
effects have negligible impact on polarimetry data, except perhaps for objects
with highly wavelength-variable polarization (potentially a concern for
SN~2002ap), and/or very blue flux.

\begin{figure}
\ssp
\begin{center}
\rotatebox{90}{
\scalebox{0.7}{
\plotone{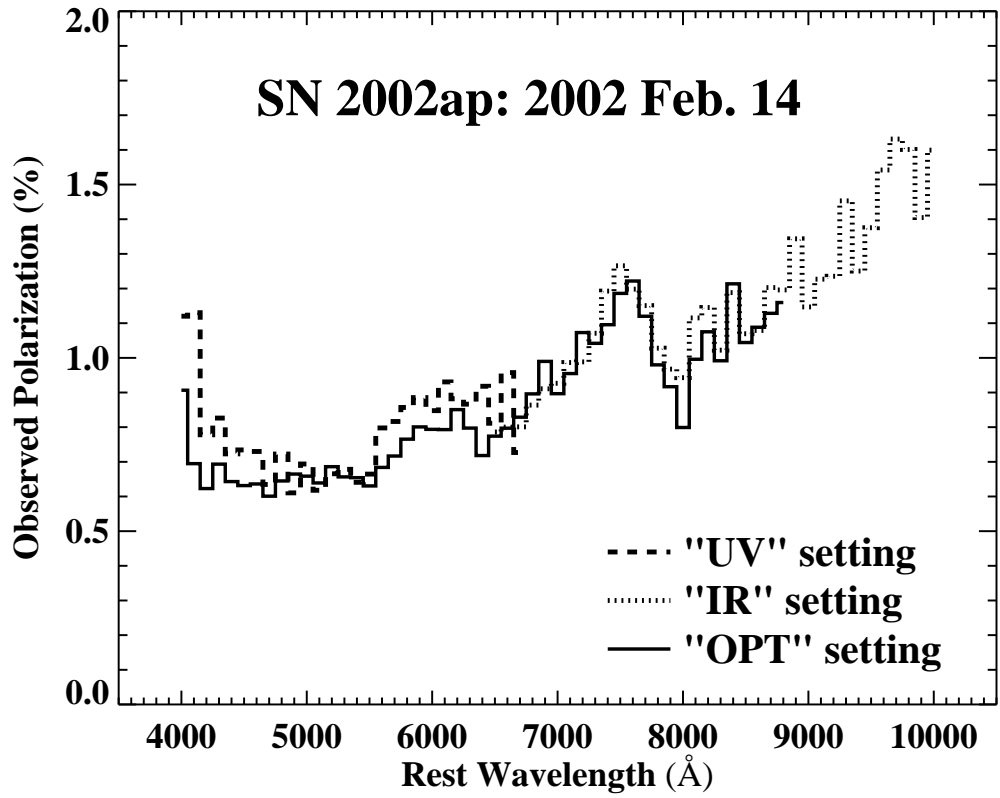}
}
}
\end{center}
\caption{Polarization of SN~2002ap obtained with the three settings described
in the text, shown at 100 \AA\ bin$^{-1}$.  UV setting data have been corrected
for the small degree of instrumental polarization described in the text.
\label{fig:10} }
\end{figure}

The redundancy of having the OPT data beyond $6400$~\AA\ also covered by the IR
setting, in which an order-blocking filter was used (in addition to the
dichroic), allows us to directly test the conclusions of Putney \& Cohen (1996)
for SN~2002ap in the Feb. 14 data.  Figure~\ref{fig:10} shows the polarimetry
data for SN~2002ap obtained on this night in all three settings.  The agreement
at red wavelengths is indeed quite good, and suggests that second-order light
effects are not a concern.  At blue wavelengths, there is relatively good
agreement between OPT and UV, although some discrepancies in the polarization
level at the $\sim 0.1\%$ level do exist.


Given the concerns with the UV data, the demonstrated lack of second-order
light effects at red wavelengths in the OPT setting, and the very limited
availability of SN~2002ap at the second epoch (it was approaching solar
conjunction and visible for less than 1 hour immediately after sunset), we
decided to observe SN~2002ap using only the OPT setting on March 7.  Thus,
direct comparison between these data and those for which an order-blocking filter
was used is not possible for this epoch.  However, we can estimate
the fraction of the observed red flux contributed by second-order blue light
through a procedure similar to that described by Putney \& Cohen (1996), and
compare this with the amount from the first epoch, for which second-order
effects were shown to be negligible.  

\begin{figure}
\ssp
\begin{center}
\rotatebox{90}{
\scalebox{0.7}{
\plotone{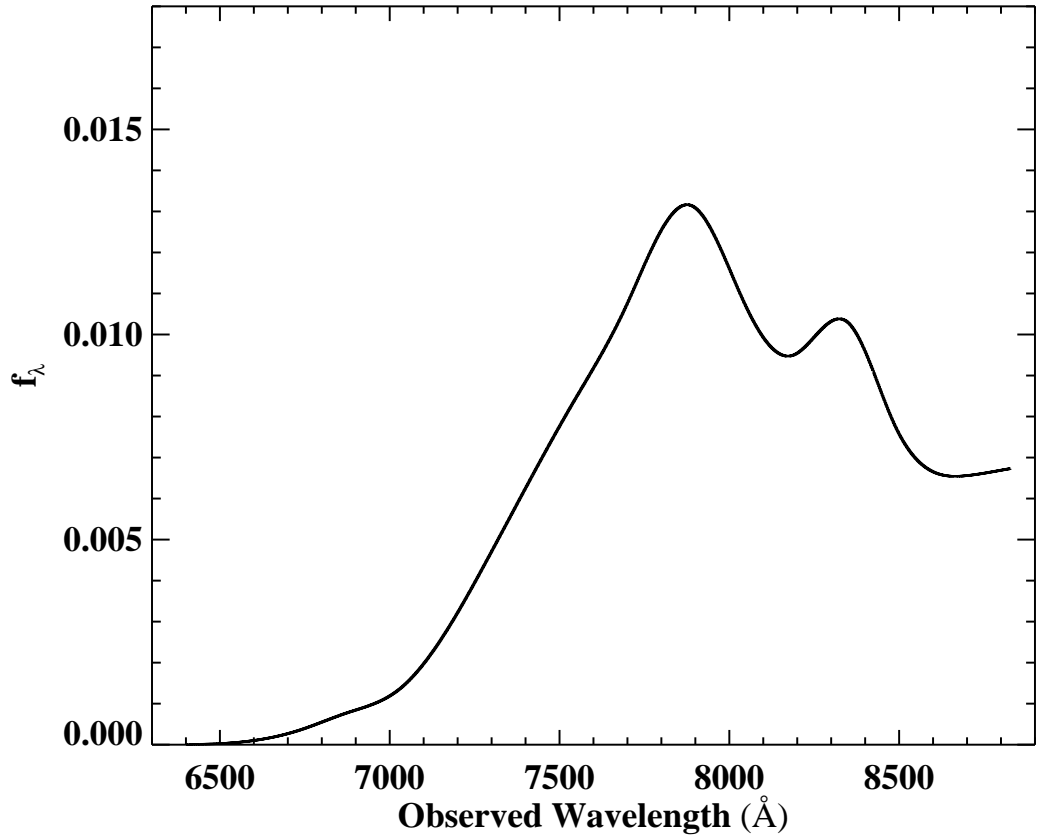}
}
}
\end{center}
\caption{Fraction of incident blue light appearing as second-order red light
calculated using equation~(\ref{eqn:a1}) and data of Feige~34 taken with LRISp.
\label{fig:11} }
\end{figure}

Since the second-order light effects derived by Putney \& Cohen (1996) were not
derived using observations made with the polarimetry optics in the light-path,
we determined our own values by observing the blue flux standard Feige 34
(Stone 1977; Massey \& Gronwall 1990) in the OPT setting, both with and without
the GG495 order-blocking filter, as well as with the UV/IR setup.  All
observations were taken with the achromatic half-wave plate at a position angle
of $0^\circ$, which produces two parallel, perpendicularly polarized beams on
the CCD chip.  The beams were individually flux calibrated and then combined to
yield the total flux spectrum.  Following the procedure of Putney \& Cohen
(1996), we calculate the fraction of blue light appearing as second-order
contamination in the red end as

\begin{equation}
f_\lambda = \frac{F^\prime_\lambda - F_\lambda}{F^\prime_{\lambda/2}},
\label{eqn:a1}
\end{equation}

\noindent where $F^\prime_\lambda$ is the flux-calibrated data measured from
6400--8828~\AA\ in the OPT setting without an order-blocking filter in place,
$F_\lambda$ is the flux-calibrated data in the same region observed with an
order-blocking filter, and $F^\prime_{\lambda/2}$ is the flux-calibrated data
measured below $4414$~\AA.  We then smoothed $f_\lambda$ and interpolated over
telluric features (e.g., the A-band) to produce the result shown in
Figure~\ref{fig:11}, which is, in fact, quite similar to the result derived by
Putney \& Cohen (1996).


\begin{figure}
\ssp
\begin{center}
\rotatebox{90}{
\scalebox{0.7}{
\plotone{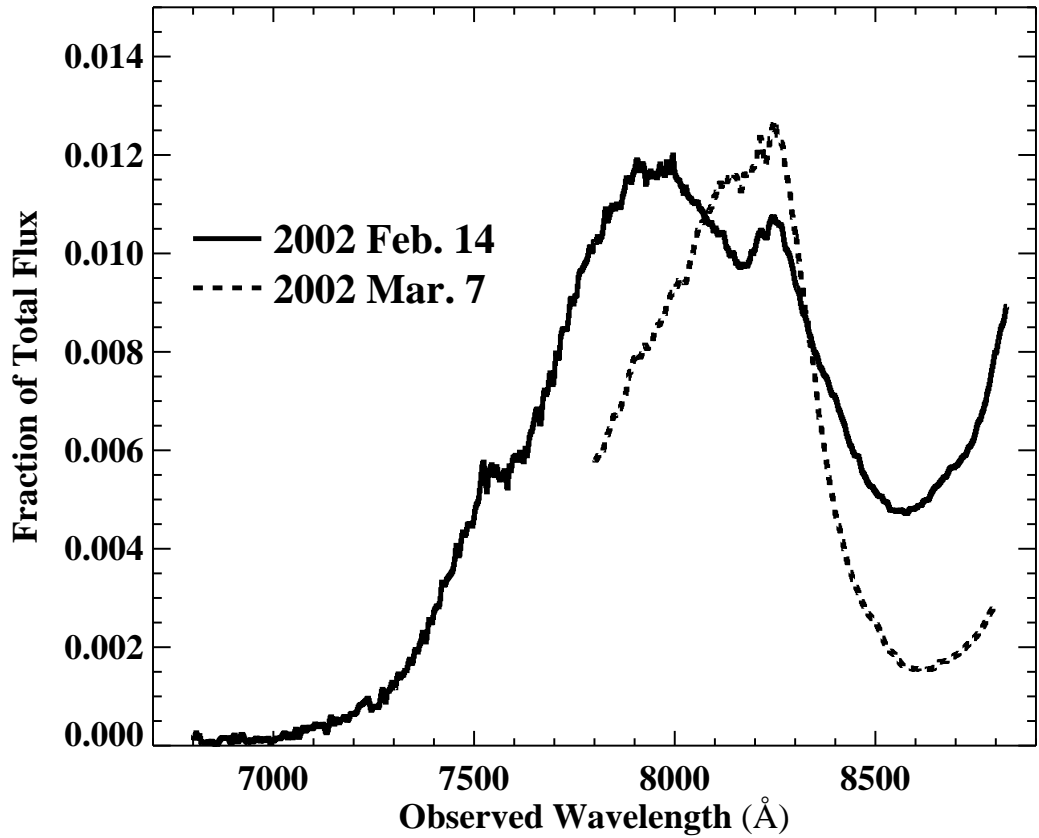}
}
}
\end{center}
\caption{Fraction of observed (total) flux contributed by second-order flux for
SN~2002ap in the OPT setting observations on 2002 Feb. 14 and Mar. 7.  
\label{fig:12} }
\end{figure}

To calculate the fraction of the observed red light contributed by second-order
blue flux for SN~2002ap, then, we calculate

\begin{equation}
\frac{F_{\rm second order}}{F_{\rm observed}} = \frac{F^\prime_{\lambda/2} \times f_\lambda}{F^\prime_\lambda},
\label{eqn:a2}
\end{equation}

\noindent for both the February and March epochs, and show the results in
Figure~\ref{fig:12}.  (Flux-calibrated data were available for SN~2002ap down to
$3400$~\AA\ on February 14 and $3900$~\AA\ on March 7; see Table~1.)  It is
clear that second-order light does not make an appreciably larger contribution
to the March 7 data than it did on February 14.  In both epoch's data, the
contribution is less than 1.3\% of the observed flux.  This similarity,
especially when coupled with the smooth, non-variable nature of the
polarization in the range $3900$--$4400$ \AA\ (Fig.~\ref{fig:2}), implies similarly
negligible second-order light effects in the March data, at least at
wavelengths beyond 7800~\AA, which includes the important \ion{Ca}{2} IR
region.

Additional confidence that second-order light does not compromise the data from
the March epoch is gained from the favorable comparison with the
spectropolarimetry data that Kawabata et al. (2002) obtained at nearly the same
epoch, presumably without any second-order light contamination.  We therefore
conclude that second-order light contamination has had a negligible impact on
our data for both epochs, and concur with the conclusions of Putney \& Cohen
(1996) that second-order light is not a concern for the spectropolarimetry of
objects that do not have either highly wavelength-variable polarization at blue
wavelengths or extremely blue color.


\end{document}